\documentclass{ws-p8-50x6-00}
\usepackage{graphics}

\def\rightnote{\em Copyright\ \copyright\ 2002 by {\bf World Scientific}, reprinted with permission.}
\def\NH3{NH$_3$}
\newcommand{\ket}[1]{\ensuremath{|#1\rangle}}
\newcommand{\bra}[1]{\ensuremath{\langle #1|}}
\newcommand{\ktau}{\ensuremath{\tilde{\kappa}({\tau})}}
\newcommand{\walk}[2]{\ensuremath{{\bf R}_{#1}^{(#2)}}}
\newcommand{\eref}{\ensuremath{E_{\mathit{ref}}}}

\title{Efficient implementation of the Projection Operator Imaginary
Time Spectral Evolution (POITSE) method for excited states}

\author{Patrick Huang, Alexandra Viel, and K.~Birgitta Whaley}

\address{Department of Chemistry and Kenneth S. Pitzer Center for
Theoretical Chemistry,\\ University of California, Berkeley, CA
94720-1460, USA}

\date{\today}
\begin{document}
\maketitle

\abstracts{We describe and systematically analyze new implementations
of the Projection Operator Imaginary Time Spectral Evolution (POITSE)
method for the Monte Carlo evaluation of excited state energies.  The
POITSE method involves the computation of a correlation function in
imaginary time.  Decay of this function contains information about
excitation energies, which can be extracted by a spectral transform.
By incorporating branching processes in the Monte Carlo propagation,
we compute these correlation functions with significantly reduced
statistical noise.  Our approach allows for the stable evaluation of
small energy differences in situations where the previous POITSE
implementation was limited by this noise.}

\section{Introduction}
\label{sec:intro}

The\footnotetext{Reprinted from P. Huang, A. Viel, and K.~B. Whaley,
in {\em Recent Advances in Quantum Monte Carlo Methods, Part II},
edited by W.~A. Lester, Jr., S.~M. Rothstein, and S. Tanaka (World
Scientific, Singapore, 2002), p.~111.} Projection Operator Imaginary
Time Spectral Evolution (POITSE\index{POITSE}) method has allowed
calculation of excited states to be made with diffusion Monte Carlo
(DMC\index{DMC}) without nodal constraints.\cite{blume97} The main
requirement is that a reasonable ground state wave function be
available, which can be obtained from well-established ground state
methods such as DMC.  The excited states are then accessed via
projector operators\index{POITSE!projector operators}, whose evolution
in imaginary time contain information on excited state energies.  In
the POITSE method a correlation function of the projection operators
is evaluated by Monte Carlo techniques, and then subsequently inverted
to obtain spectral functions whose peak positions correspond to
excited state energies.  This inversion requires an inverse Laplace
transform\index{Laplace transform, inverse}, a notoriously
ill-conditioned numerical procedure.  In the applications of POITSE
made to date,\cite{blume97b,blume99,blume00b} this inversion has been
performed with the Maximum Entropy Method
(MEM\index{MEM}).\cite{bryan90} POITSE has considerable power in
allowing analysis of excited states without imposing nodal
restrictions.  It is particularly useful when some physical insight
about the nature of the desired excited states is available.  This
information can be used to tailor suitable projectors to obtain
maximum overlap with the eigenstates of interest.  This has been
demonstrated recently with permutation symmetry tunneling
excitations.\cite{blume00b} In general, the viability and power of the
method has now been shown for a range of model systems involving
atomic motions.\cite{blume97b} It has been applied to several physical
examples of cluster excitations which cannot be addressed by basis set
methods, including up to 15-dimensional problems.\cite{blume99} To our
knowledge, the method has not yet been systematically applied to
fermion problems, although there is no intrinsic impediment to this.

In this paper we analyze the efficiency and accuracy of the POITSE
algorithm for various different implementations of the DMC component
of the method.  We present a modification of the algorithm that allows
the calculation of small energy differences with reduced statistical
noise.  In Sec.~\ref{sec:methods}, we briefly review the POITSE
general formalism and explain in detail the different numerical
implementations.  Sec.~\ref{sec:examples} illustrates the different
implementations with two applications: the one-dimensional problem of
the ammonia inversion mode and the six-dimensional van der Waals
vibration of the $^4$He-benzene dimer.

\section{Computational Methodology}
\label{sec:methods}

The general POITSE method involves the Monte Carlo evaluation of an
imaginary time ($\tau=it$) correlation function $\ktau$, and then a
subsequent inverse Laplace transform of this correlation function
using the Maximum Entropy Method\index{MEM}.  With an appropriately
chosen correlation function, the inverse Laplace transform provides a
spectral function whose peak positions correspond to excitation
energies.  The basic theory\cite{blume97} and its application to model
systems\cite{blume97b,blume99,blume00b} have previously been described
in detail, and thus we will only present a brief summary of the
relevant formalism.

\subsection{Theory}
\label{subsec:theory}

The\index{POITSE!theory|(} primary quantity of interest in POITSE is
the spectral function $\kappa(E)$,
\begin{equation}
\kappa(E) = \sum_n |\bra{\phi_0}\hat{A}\ket{\phi_n}|^2 \delta(E-E_n+E_0),
\label{eq:spec}
\end{equation}
where $\{\ket{\phi_n}\}$ and $\{E_n\}$ are a complete set of energy
eigenkets and eigenenergies for the Hamiltonian $\hat{H}$, and
$\hat{A}$ is an operator chosen to connect $\ket{\phi_0}$ at least
approximately to the particular excited state(s) of interest
$\ket{\phi_n}$.  Taking the Laplace transform of Eq.~(\ref{eq:spec}),
one can obtain the imaginary time correlation function $\ktau$, in
atomic units ($\hbar=1$), as
\begin{eqnarray}
\ktau & = & \bra{\phi_0} \hat{A} e^{-(\hat{H}-E_0)\tau} \hat{A}^{\dagger}
\ket{\phi_0} \label{eq:corr}\\
      & = & \sum_n |\bra{\phi_0}\hat{A}\ket{\phi_n}|^2 e^{-(E_n-E_0)\tau}.
\label{eq:decay}
\end{eqnarray}
The POITSE approach consists of evaluating $\ktau$ by a Monte Carlo
algorithm, then taking its inverse Laplace transform\index{Laplace
transform, inverse} to obtain the spectral function $\kappa(E)$.

In most situations, however, the ground state $\ket{\phi_0}$ is not
known exactly.  In practice, one typically employs a trial
function\index{trial functions} $\ket{\Psi_T}$ and reference energy
$\eref$ which approximate as closely as possible $\ket{\phi_0}$ and
$E_0$, respectively.  Use of a reference energy not equal to the exact
ground state energy modifies the decay rate of all terms in
Eq.~(\ref{eq:decay}) by a constant factor of $\eref-E_0$.  This
results in a systematic bias in the excitation energies of
Eq.~(\ref{eq:spec}), independent of the usual finite time step
bias\index{DMC!time step bias} due to the DMC evaluation of
Eq.~(\ref{eq:decay}).  This bias from $\eref$ is also independent of
whether the true ground state $\ket{\phi_0}$ is used.

It has been shown earlier\cite{blume97} that such systematic bias can
be eliminated by introducing the normalization factor
\begin{equation}
\bra{\Psi_T} e^{-(\hat{H}-\eref)\tau} \ket{\Psi_T}. \label{eq:norm}
\end{equation}
The removal of bias due to $\eref$ can be seen from the following
arguments.  First, replacing $\ket{\phi_0},E_0$ in Eq.~(\ref{eq:corr})
with $\ket{\Psi_T},\eref$, respectively, and dividing by the
additional normalization factor of Eq.~(\ref{eq:norm}), leads to the
modified decay function
\begin{equation}
\ktau = \frac{\bra{\Psi_T} \hat{A} e^{-(\hat{H}-\eref)\tau}
\hat{A}^{\dagger} \ket{\Psi_T}}{\bra{\Psi_T} e^{-(\hat{H}-\eref)\tau}
\ket{\Psi_T}}. \label{eq:corr2}
\end{equation}
$\ket{\Psi_T}$ is then expanded in eigenstates of $\hat{H}$ to
yield\cite{blume97,blume98}
\begin{equation}
\ktau = \frac{\sum_n |\bra{\Psi_T}\hat{A}\ket{\phi_n}|^2
e^{-(E_n-\eref)\tau}}{\sum_m c_m^2 e^{-(E_m-\eref)\tau}}, \label{eq:corr2b}
\end{equation}
where $c_m=\bra{\Psi_T}\phi_m\rangle$.  The numerator and denominator
of Eq.~(\ref{eq:corr2b}) may then be multiplied by
$e^{(E_0-\eref)\tau}/c_0^2$ to obtain
\begin{eqnarray}
\ktau & = & \left[1+\sum_{m=1} \left(\frac{c_m}{c_0}\right)^2
e^{-(E_m-E_0)\tau}\right]^{-1} \sum_n
\left|\frac{\bra{\Psi_T}\hat{A}\ket{\phi_n}}{\bra{\Psi_T}\phi_0\rangle}\right|^2 e^{-(E_n-E_0)\tau} \label{eq:corr2c} \\
      & \propto & \sum_n |\bra{\Psi_T}\hat{A}\ket{\phi_n}|^2
e^{-(E_n-E_0)\tau} + O(x). \label{eq:corr2d}
\end{eqnarray}
Here, the prefactor of Eq.~(\ref{eq:corr2c}) was expanded in a power
series in $x$, where
\begin{equation}
x = \sum_{m=1} \left(\frac{c_m}{c_0}\right)^2 e^{-(E_m-E_0)\tau}.
\end{equation}
When $\ket{\Psi_T} = \ket{\phi_0}$, we see that Eq.~(\ref{eq:corr2d})
is identically equal to Eq.~(\ref{eq:decay}), and the effects of using
a reference energy other than the true ground state energy are
completely eliminated.  Additive errors of $O(x)$ and higher are
present when an approximate ground state is used.  Note that since the
series expansion of Eq.~(\ref{eq:corr2c}) is only convergent for
$c_0>\sqrt\frac{1}{2}$, this does require that a reasonable
approximation to the ground state be available.  The higher order
terms $O(x)$ contribute to the spectral function $\kappa(E)$ in an
additive manner.  Consequently, they do not affect the positions of
the relevant spectral features of interest, {\em i.e.} the dominant
leading terms of Eq.~(\ref{eq:corr2d}).  In practice, for a reasonable
choice of $\ket{\Psi_T}$ these additional terms have highly reduced
weight.\cite{blume97,blume98} To leading order therefore, the
renormalized decay Eq.~(\ref{eq:corr2}) exhibits the time dependence
of Eq.~(\ref{eq:decay}), independent of the reference energy $\eref$.
Consequently $\eref$ may be arbitrarily chosen and varied.  The
usefulness of this will become more apparent below, in our discussion
of numerical implementation.

The numerical inversion of $\ktau$ to obtain $\kappa(E)$ is an
ill-conditioned problem, especially when Monte Carlo noise is
non-negligible and/or when the spectral function $\kappa(E)$ contains
multiple overlapping peaks of comparable intensity.  Thus a judicious
choice of the operator $\hat{A}$ is necessary to ensure that the
time-dependence of $\ktau$ is dominated by only one or a few
well-separated energy differences.  The inverse Laplace transform of
$\ktau$ is performed using the Bryan implementation\cite{bryan90} of
the maximum entropy\index{MEM} method.  Our use of this approach for
the inversion of $\ktau$ is identical to that employed in previous
POITSE work.\cite{blume97,blume99,blume00b,blume97b} We will discuss
choices for $\hat{A}$ specific to particular systems of study in
Sec.~\ref{sec:examples}.\index{POITSE!theory|)}

\subsection{Numerical Implementation}
\label{subsec:implement}

The\index{POITSE!implementation|(} correlation function of
Eq.~(\ref{eq:corr2}) may be rewritten in a form amenable to Monte
Carlo evaluation as\cite{blume97}
\begin{equation}
\ktau = \frac{\sum_j \hat{A}^{\dagger}(\walk{j}{0}) \hat{A}(\walk{j}{\tau})
w(\walk{j}{\tau})}{\sum_j w(\walk{j}{\tau})}, \label{eq:ktaumc}
\end{equation}
where $\walk{j}{\tau}$ is a guided random walk $j$ in multidimensional
configuration space, discretized in time steps of size $\Delta\tau$ (a
DMC ``walker''), and
\begin{eqnarray}
w(\walk{j}{\tau}) & = & \prod_m
\exp{\{-[E_L(\walk{j}{m\Delta\tau})-\eref]\Delta\tau\}}, \\
E_L(\walk{j}{\tau}) & = &
\Psi_T^{-1}(\walk{j}{\tau})\hat{H}\Psi_T(\walk{j}{\tau}).
\end{eqnarray}
The quantities $w(\walk{j}{\tau})$ and $E_L(\walk{j}{\tau})$ are the
usual DMC cumulative weight and local energy,
respectively.\cite{hammond94} The evaluation of Eq.~(\ref{eq:ktaumc})
begins with a variational Monte Carlo (VMC)\index{VMC} walk in which
an initial starting ensemble of walkers distributed according to
$\Psi_T^2({\bf R})$ is generated using a simple Metropolis
method.\cite{hammond94} The starting VMC ensemble is subsequently
propagated in imaginary time by a DMC sidewalk\index{DMC}, during
which Eq.~(\ref{eq:ktaumc}) is sampled.  Since the maximum entropy
analysis requires independent samples of $\ktau$, the starting
configuration for each DMC sidewalk is taken from the VMC walk every
$100-200$ VMC steps apart, to minimize correlations between successive
sidewalks.  The set of $\ktau$'s evaluated in this manner serve as
input for the inverse Laplace transform via MEM\index{MEM}.  Typically
$100-500$ independent decays are required to produce a converged
spectrum $\kappa(E)$.

In the original implementation of Blume~{\em et al.},\cite{blume97}
the DMC weights $w(\walk{j}{\tau})$ take on a continuous range of
values, and walkers are not destroyed or duplicated.  We refer to this
approach here as DMC with pure weights\index{DMC!pure weighting}.
This is the preferable implementation in an ideal situation where
high-quality trial functions are available.  However, for reasonably
complex systems this is often not the case.  In addition, it has been
shown that DMC with pure weights is unstable for long propagation
times.\cite{assaraf00} Therefore, as we demonstrate in
Sec.~\ref{sec:examples}, a DMC sidewalk that uses pure weights may
sometimes be impractical in situations involving small energy
differences.

A common solution to the problems associated with pure weights is to
introduce branching\index{DMC!pure branching}.  The simplest
branching scheme rounds the walker weight at every step of the walk to
an integer $n_j = \mathrm{int}[w(\walk{j}{\tau})+\xi]$, where $\xi$ is
an uniformly distributed random number on $[0,1)$.  A walker
$\walk{j}{\tau}$ is destroyed for $n_j=0$; otherwise, $n_j$ copies of
walker $\walk{j}{\tau}$ are propagated independently in the next DMC
move.  In this case, the weights $w(\walk{j}{\tau})$ take on only
integer values, and Eq.~(\ref{eq:ktaumc}) becomes
\begin{equation}
\ktau = \frac{1}{n_w} \sum_{j'}^{n_w} \hat{A}^{\dagger}(\walk{j}{0})
\hat{A}(\walk{j'}{\tau}),
\end{equation}
where the index $j$ denotes the parent walker at initial time $\tau=0$
from which walker $j'$ at time $\tau$ descended, and the instantaneous
ensemble size $n_w$ fluctuates with time.  We refer to this approach
here as DMC with pure branching.  While the pure branching method is
formally correct on average and is much more stable numerically, the
integer rounding of walker weights can nevertheless lead to greater
statistical noise.\cite{barnett91}

To minimize this noise, one can employ a hybrid
approach\index{DMC!hybrid branching/weighting} where each weight
$w(\walk{j}{\tau})$ is allowed to vary continuously, and a walker is
only destroyed or duplicated when its weight exceeds some
predetermined bounds.  In such a situation, it is important that the
branching procedure does not artificially alter the ensemble sum of
weights $W_{\mathit{tot}}=\sum_j w(\walk{j}{\tau})$.  A combined
weighting and branching scheme will in general exhibit less
statistical noise than a pure branching scheme.  In some cases the
noise reduction can be significant.  Our implementation of branching
is similar to that outlined in Ref.~\ref{blume96}.  About every
$20-50$ DMC time steps, the ensemble is checked for walkers whose
weight exceeds the empirically determined bounds $w_{\mathit{min}}$
and $w_{\mathit{max}}$.  A walker $\walk{j}{\tau}$ with weight
$w(\walk{j}{\tau})>w_{\mathit{max}}$ is split into $n_j =
\mathrm{int}[w(\walk{j}{\tau})+\xi]$ walkers, each with weight
$w(\walk{j}{\tau})/n_j$.  A walker $\walk{j}{\tau}$ with weight
$w(\walk{j}{\tau})<w_{\mathit{min}}$ is either a) killed with
probability $1-w(\walk{j}{\tau})$, otherwise b) kept with its weight
set to unity.  The bounds $w_{\mathit{min}}$ and $w_{\mathit{max}}$
are chosen to give a stable DMC walk with respect to the ensemble size
and $W_{\mathit{tot}}$.

As discussed previously, incorporation of the normalization factor of
Eq.~(\ref{eq:norm}) into $\ktau$ results in a decay independent of the
reference energy $\eref$.  Therefore we are free to choose and vary
$\eref$ based on considerations of numerical stability.  A common
choice of $\eref$ is the variational energy of the trial function,
$\eref=\bra{\Psi_T}\hat{H}\ket{\Psi_T}/\bra{\Psi_T}\Psi_T\rangle$,
which may be obtained from a separate VMC calculation.  One may also
choose the ground state energy $\eref=E_0$, which is readily obtained
from standard ground state DMC methods.  In our implementation, we
begin with an initial choice of $\eref$ and update $\eref$
continuously during the course of the DMC walk according to
\begin{equation}
\eref^{(\tau+\Delta\tau)} = \eref^{(\tau)} +
\frac{\eta}{\Delta\tau}\ln{\left[\frac{\sum_j w({\bf R}_j^{(\tau)})}{\sum_j
w({\bf R}_j^{(\tau+\Delta\tau)})}\right]} \label{eq:eref}
\end{equation}
where $\eta$ is an empirical update parameter chosen to be as small as
possible to avoid biasing the results, typically
$\eta/\Delta\tau=0.01-0.3$.  The effect of this updating procedure for
$\eref$ is to keep the average walker weight close to unity, thus
preventing the ensemble size and sum of weights from diverging off to
infinity or zero.  The combination of these various mechanisms serve
to ensure a stable DMC walk for long times, thus allowing the
evaluation of small energy differences $E_n-E_0$.  In the examples
presented in Sec.~\ref{sec:examples} we will compare the effects of
the various DMC schemes described here.

A final note in the implementation concerns the statistical
errors\index{MEM!errors} in the excited state energy differences
$E_n-E_0$.  The MEM\index{MEM} inversion of $\ktau$ gives the spectral
function $\kappa(E)$, whose peak positions correspond to excited state
energy differences.  There is no general approach to assign error bars
in the mean peak position,\cite{blume97b} and thus we only report
energies to the last significant figure.  We determine empirically the
position of this last significant figure by examining the convergence
of $E_n-E_0$ with respect to the number of decays $\ktau$ used as
input for the MEM inversion.  Because multiple projectors are usually
sampled from the same DMC sidewalk, the {\em relative} differences
between excited states are expected to be very
accurate.\index{POITSE!implementation|)}

\section{Examples}
\label{sec:examples}

\subsection{\NH3 inversion}
The first application we discuss here is a POITSE study of the ammonia
inversion mode.  Freezing all other internal degrees of freedom, the
Schr\"odinger equation for this mode alone is a one-dimensional
problem which can be solved exactly by a straightforward Discrete
Variable Representation--Finite Basis Representation
(DVR--FBR)\index{DVR} calculation.\cite{leforestier91} The Hamiltonian
is given by
\begin{equation}
\hat H = -\frac{\hbar^2}{2\mu}\frac{\partial^2}{\partial h^2} + V(h),
\end{equation}
where $h$ is the distance between the nitrogen atom and the hydrogen
plane, $\mu$ is the effective mass for the mode and $V(h)$ is the
double-well inversion potential for tunneling\index{tunneling!ammonia
inversion} across the hydrogen plane.  We use one of the potential
forms (``Case b'') proposed by Ni\~no {\it et al.}\cite{nino95} which
leads to a tunneling splitting of 1.43~cm$^{-1}$ for the lowest
tunneling pair, and 64.5~cm$^{-1}$ for the next lowest tunneling pair.
The corresponding DVR--FBR energy levels\index{DVR} are listed in
Table~\ref{table:nh3} as benchmarks for the POITSE results.
\begin{table}[b]
\caption[]{Lowest four energy levels (in~cm$^{-1}$) for the inversion
mode of \NH3 relative to the ground state energy, which is
553.11~cm$^{-1}$ above the potential minimum.}
\begin{center}
\begin{tabular}{|c|cccc|}
\hline
&		$E_0$ &	$E_1$ &	$E_2$ &	$E_3$	\\
\hline
DVR--FBR &	0.00 &	1.43 &	961.40 & 1025.93 \\
\hline
\end{tabular}
\end{center}
\label{table:nh3}
\end{table}

A double well study was previously made in Ref.~\ref{blume97b} to
demonstrate the effectiveness of the POITSE method for model systems.
However, in that example, the energy differences involved were much
larger than those arising in the \NH3 inversion problem which we
discuss here.  While the inversion frequency is high (993~cm$^{-1}$),
the POITSE method allows the computation of an energy difference which
is three orders of magnitude smaller.  We compare here two different
DMC implementations, namely pure weights\index{DMC!pure weighting} and
pure branching\index{DMC!pure branching}, and demonstrate the
limitations associated with the former approach for computing small
energy differences.  In order to make such a comparison of
implementations, it is convenient and indeed preferable to use a
system for which exact wave functions can be found.

The trial function\index{trial functions!ammonia inversion}
$\Psi_T(h)$ used in the Monte Carlo evaluation of Eq.~(\ref{eq:corr2})
was initially fit to the DVR--FBR ground state
eigenfunction\index{DVR} $\Phi_0(h)$, and then further optimized by
VMC.\index{VMC!optimization} While numerous sophisticated VMC
optimization schemes exist,\cite{umrigar88,snajdr00,nightingale01} for
a simple one-dimensional problem we found it sufficient to manually
vary the trial function parameters to minimize the ground state energy
and its variance.  We use the analytical form
\begin{equation}
\Psi_T(h) = \exp [ a_0e^{b_0(h-c_0)^2} + a_0e^{b_0(h+c_0)^2} +
d_0e^{e_0h^4} ],
\end{equation}
where $a_0,b_0,c_0,d_0,$ and $e_0$ are parameters listed in
Table~\ref{table:NH3_para}.  The corresponding VMC energy is
561.4(3)~cm$^{-1}$, which is less than 2\% above the exact ground
state value obtained from DVR--FBR.

Since the first excited state $\Phi_1(h)$ of a double well potential
is the lowest antisymmetric state, the
projector\index{POITSE!projector operators|(} $\hat{A}=h$ was
previously used\cite{blume97b} to access this level.  In obtaining
higher-lying states, choosing $\hat{A}$ to be an integer power of $h$
led to a $\ktau$ consisting of a superposition of multiple exponential
decays.  For instance, a choice of $\hat{A}\Psi_T(h) = h^2\Psi_T(h)$
resulted in non-negligible overlap with multiple excited levels.  Thus
an accurate Laplace inversion of the corresponding $\ktau$ was more
difficult, due to the multiple decay contributions of these states.
We use here instead more effective projectors given by the ratio of
the eigenfunctions
\begin{equation}
\hat{A}_n = \frac{\Phi_n(h)}{\Psi_T(h)}, \label{eq:nh3_proj}
\end{equation}
where $\Phi_n(h)$ is an excited state eigenfunction obtained from a
DVR--FBR calculation.\index{DVR} Clearly if the eigenfunctions are
numerically exact, this results in an exact projector.  Such
projectors have also been shown to be useful when only symmetry
properties of the eigenfunctions are well
characterized.\cite{blume00b} The following analytical expressions
were fitted to the DVR--FBR eigenfunctions for the lowest three
excited states ($n=1-3$):
\begin{eqnarray}
\Phi_1(h) & = &	e^{b_1(h-c_1)^2} - e^{b_1(h+c_1)^2} \\
\Phi_2(h) & = &	a_2[(h-f_2)e^{b_2(h-c_2)^2} - (h+f_2)e^{b_2(h+c_2)^2}] +
d_2e^{e_2h^4} \\
\Phi_3(h) & = &	a_3[(h-f_3)e^{b_3(h-c_3)^2} + (h+f_3)e^{b_3(h+c_3)^2}] +
d_3he^{e_3h^4}.
\end{eqnarray}
The fit parameters are given in Table~\ref{table:NH3_para}.  We
emphasize that we are using this example of ammonia inversion to
demonstrate and compare the relative efficiency of two different and
alternative implementations of the POITSE algorithm.  In particular,
we shall compare the extent of noise and time step bias of the two
different approaches to the DMC evaluation of Eq.~(\ref{eq:corr2}).
Our aim here is not the establish the generality of the method, or its
accuracy for a double well problem, both of which have been addressed
in earlier work.\cite{blume97,blume97b} Instead, we are interested in
assessing the relative efficiency of different implementations, and
thus it is preferable here to use projectors which are as exact as
possible.\index{POITSE!projector operators|)}
\begin{table}[t]
\caption[]{Fit parameters (in atomic units) for \NH3 eigenfunctions
obtained from DVR--FBR.}
\begin{center}
\begin{tabular}{|ccccccc|}
\hline
$n$ & $a_n$ & $b_n$   & $c_n$ & $d_n$   & $e_n$  & $f_n$ \\
\hline
0 & 17.0  & -1.095  & 0.829 & 1.      & -0.054 & \\
1 &       & -10.886 & 0.681 &         &        & \\
2 & 0.785 & -11.072 & 0.674 & -0.0664 & -0.082 & 0.580 \\
3 & 0.768 & -12.025 & 0.720 & -0.314  & -1.325 & 0.447 \\
\hline
\end{tabular}
\label{table:NH3_para}
\end{center}
\end{table}

\begin{figure}[b]
\begin{center}
\scalebox{0.32}{\includegraphics{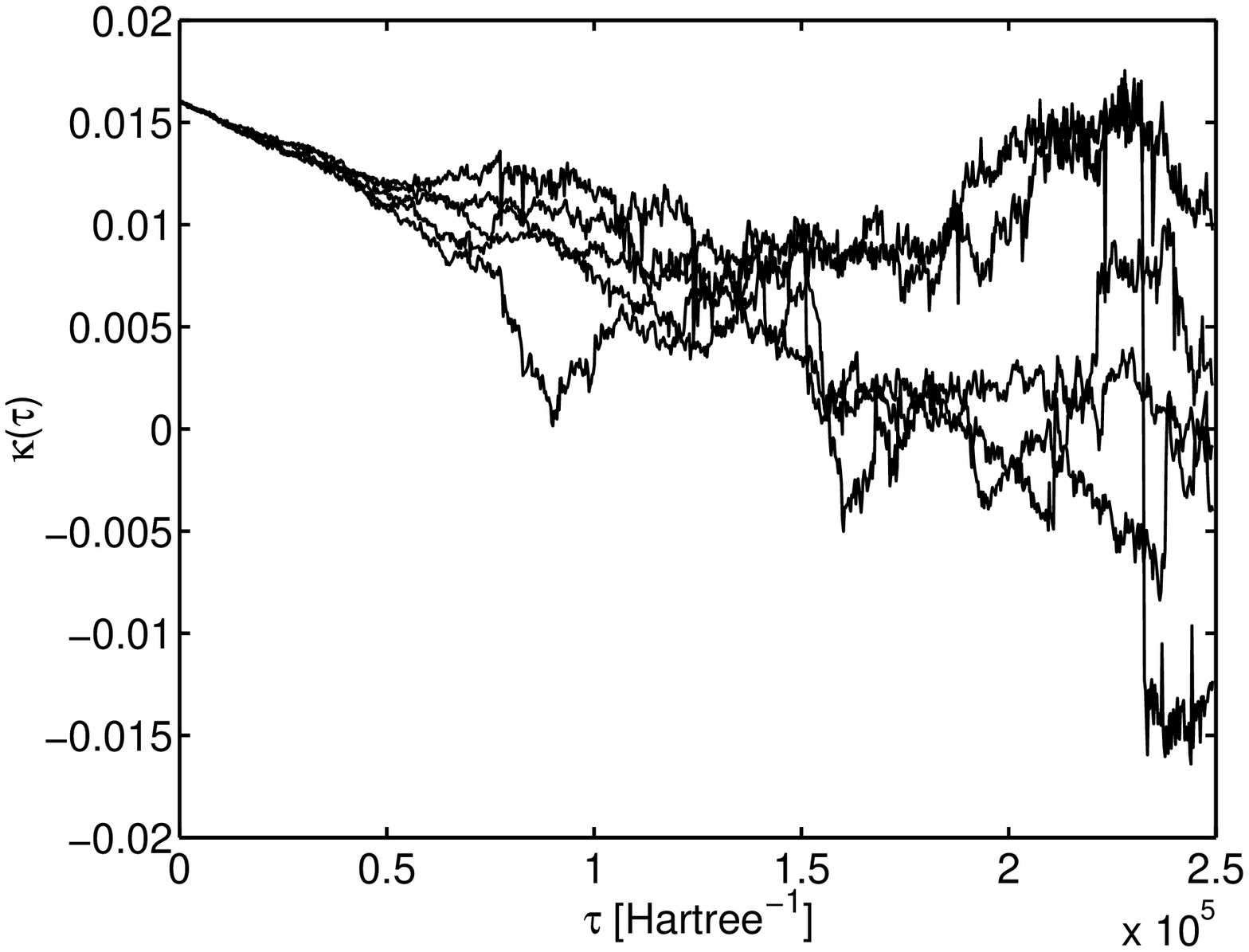}}
\scalebox{0.32}{\includegraphics{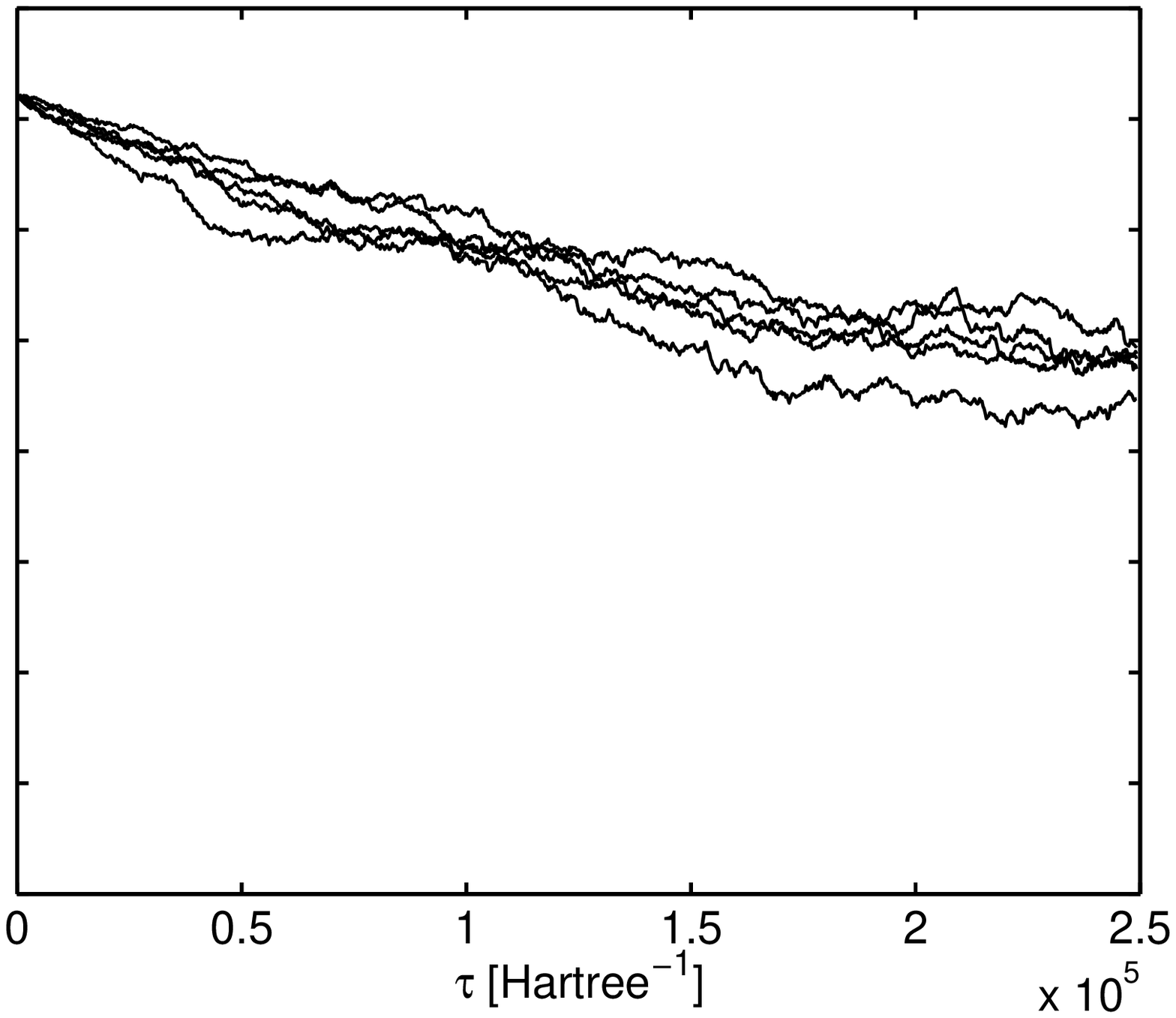}}
\end{center}
\caption[]{Typical correlation functions $\ktau$ for \NH3 using the
projector $\Phi_1/\Psi_T$.  The left plot (a) corresponds $\ktau$
evaluated using DMC with pure weights,\index{DMC!pure weighting} while
the decay curves in the right plot (b) are obtained using DMC with
pure branching.\index{DMC!pure branching}}
\label{fig:NH3_decay}
\end{figure}
Since the lowest tunneling splitting\index{tunneling!ammonia
inversion} $E_1-E_0$ is small, the corresponding decay $\ktau$ is slow
and requires a long DMC propagation.  Fig.~\ref{fig:NH3_decay}a shows
four typical $\ktau$'s computed using the original POITSE
implementation involving DMC with pure weights.\index{DMC!pure
weights} These decays become extremely noisy as the time $\tau$
increases.  The ensemble local energy $\langle E_L\rangle$ also
exhibits such behavior.  This problem is well-known\cite{assaraf00}
and arises from the fact that for long or even moderate DMC
propagation times, the Monte Carlo ensemble averages are dominated by
only a few walkers carrying high relative weights.  In comparison,
Fig.~\ref{fig:NH3_decay}b shows typical $\ktau$'s obtained from an
implementation using DMC sidewalks with pure branching\index{DMC!pure
branching}, where walkers are replicated or destroyed at each time
step based on integer rounding of their weights as discussed in
Sec.~\ref{subsec:implement}.  In both calculations, 2000 walkers were
propagated using a time step $\Delta\tau$ of 5~Hartree$^{-1}$.
Clearly there is far less noise at longer times in the pure branching
implementation, and thus such an approach is more suitable for the
evaluation of small energy differences.  Using the pure branching
scheme, the Laplace inversion of 600 decays computed up to a final
time $\tau_f$ of 250000~Hartree$^{-1}$ results in a single peak at
1.39~cm$^{-1}$, in reasonable agreement with the DVR--FBR value.

The evaluation of the larger energy differences $E_2-E_0$ and
$E_3-E_0$ are manageable using both DMC implementations, because the
lengths of the corresponding decays are much shorter than for the
lowest energy difference $E_1-E_0$.  The use of the projector given in
Eq.~(\ref{eq:nh3_proj}) facilitates the Laplace inversion, since each
choice of $\hat{A}_n$ results in a $\ktau$ consisting of only one
exponential decay.  In these calculations, 1000 decays are used as
input for the MEM inversion, with each decay computed using an
ensemble of 2000 DMC walkers propagated to a final time $\tau_f$ of
1500~Hartree$^{-1}$.  The number of decays required for a converged
$\kappa(E)$ depends on the energy difference of interest and on the
time step $\Delta\tau$.  In general, for larger time steps, DMC with
pure weights requires more sampling to produce fully converged
results.

\begin{figure}[t]
\begin{center}
\scalebox{0.32}{\includegraphics{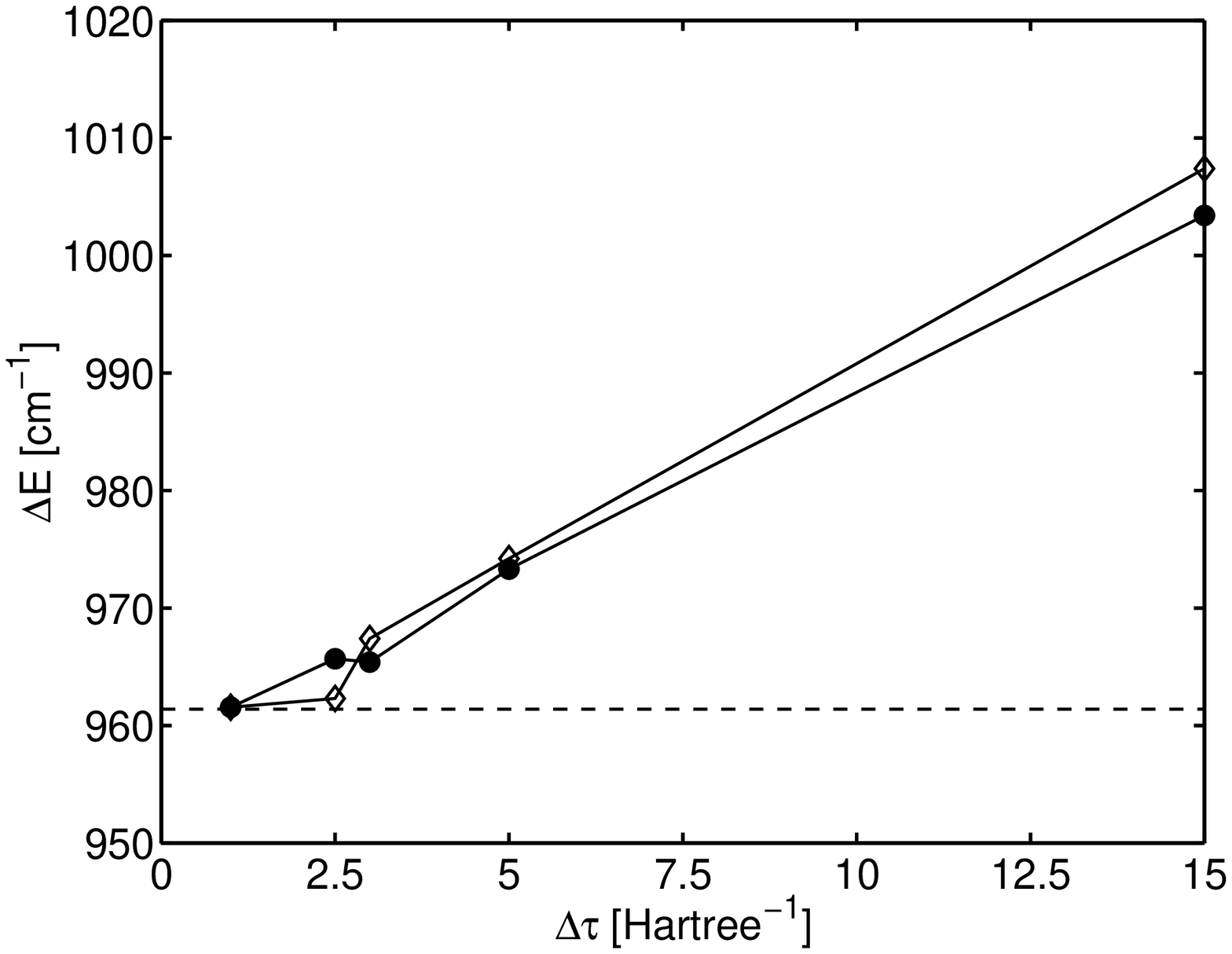}}
\scalebox{0.32}{\includegraphics{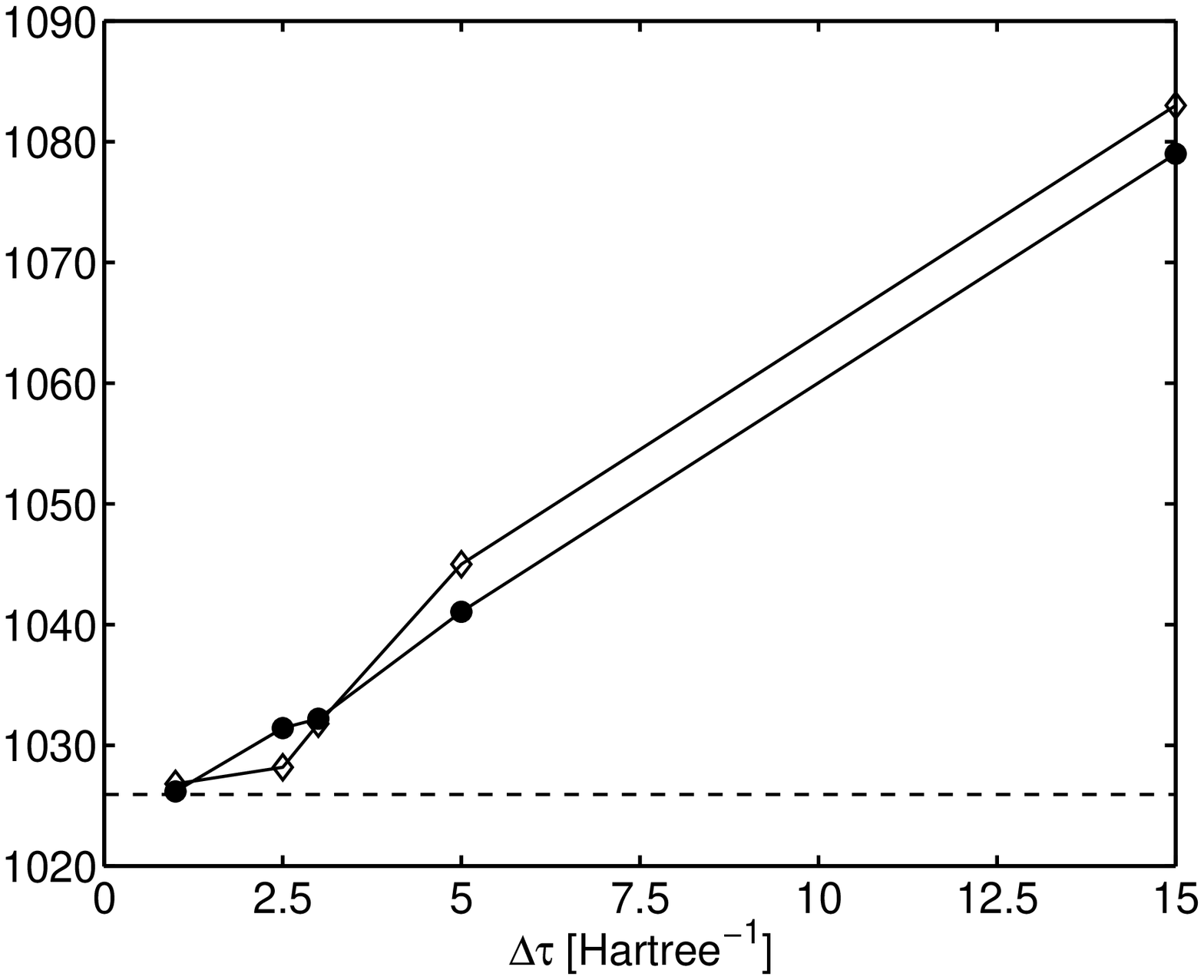}}
\end{center}
\caption[]{Time step dependence for the energy differences $E_2-E_0$
(left) and $E_3-E_0$ (right) of \NH3 inversion mode.  The dashed lines
correspond to the exact DVR--FBR values.  Energies obtained from DMC
with pure weights\index{DMC!pure weighting} are marked with filled
circles, and energies obtained from DMC with pure
branching\index{DMC!pure branching} are marked with open diamonds.}
\label{fig:E_dt}
\end{figure}

Since DMC methods are subject to a systematic time step
bias,\index{DMC!time step bias} we perform a comparative study of the
two implementations and their time step dependence.  For the
computation of the lowest energy difference $E_1-E_0$ using DMC with
pure branching, we find a time step of 5~Hartree$^{-1}$ to be
sufficiently small to give an accurate result within statistical
error.  However, the time step dependence of higher energy differences
is not necessarily the same as that for $E_1-E_0$.
Fig.~\ref{fig:E_dt} presents the time step dependence for the
calculation of $E_2-E_0$ and $E_3-E_0$, using both DMC with pure
weights\index{DMC!pure weights} (solid circles) and DMC with pure
branching (open diamonds).  It is evident that for both DMC
implementations, the higher energy differences are more sensitive to
time step bias than the lowest energy difference, $E_1-E_0$.  Thus, in
order to extract the correct energies in the higher energy range,
either a smaller time step would need to be used, or an extrapolation
to $\Delta\tau=0$ would need to be performed.

With this simple example, we have shown that two different POITSE
implementations, namely DMC with pure weights and DMC with pure
branching, lead to the same results.  We have also presented a
systematic study of the convergence behavior for these two different
approaches, and compared with the exact solution obtained from
DVR--FBR calculations.  For the evaluation of small energy
differences, we conclude that a pure branching DMC sidewalk is
considerably more efficient than using DMC with pure weights.

\subsection{$^4$He-benzene dimer}

We now demonstrate the use of the POITSE approach for the computation
of excited vibrational energies of the $^4$He-benzene dimer.  We treat
the benzene as a rigid molecule, and for simplicity we also neglect
the rotational kinetic energy of the benzene, {\em i.e.} the rotation
of benzene relative to helium.  In the space-fixed frame, the
resulting Hamiltonian is
\begin{equation}
\hat{H} = -\frac{\hbar^2}{2m_0}\nabla_0^2 - \frac{\hbar^2}{2m}\nabla_k^2 +
V({\bf r}), \label{eq:ben_ham}
\end{equation}
where $m_0$ is the benzene mass, $m$ is the helium mass, $\nabla_0^2$
is the Laplacian with respect to the benzene center-of-mass position
${\bf r}_0$, $\nabla_k^2$ is the Laplacian with respect to the helium
position ${\bf r}_k$, and $V({\bf r})$ is the $^4$He-benzene
interaction potential.  The latter depends only on the relative
coordinate vector ${\bf r}={\bf r}_k-{\bf r}_0$.  The potential is an
analytical fit\cite{kwon01} to {\em ab initio} MP2 calculations of
Hobza {\em et al.},\cite{hobza92} and possesses two equivalent global
minima of $-66.01$~cm$^{-1}$ along the six-fold $C_6$-axis, situated
at 3.27~\AA\ above and below the benzene plane.  While in principle
one could transform the Hamiltonian to the center-of-mass frame to
yield a three-dimensional problem, as would typically be done in a
basis set calculation, sampling the transformed kinetic energy terms
becomes more complicated in DMC as additional particles are added, and
thus it is technically simpler for us to work with the six-dimensional
Hamiltonian as written in Eq.~(\ref{eq:ben_ham}).

The trial function\index{trial functions!$^4$He-benzene} $\Psi_T({\bf
r})$ is the product of an anisotropic Gaussian binding factor centered
on the benzene center-of-mass, and an atom-atom repulsive factor,
\begin{equation}
\Psi_T({\bf r}) = e^{-a(x^2+y^2)-cz^2} \prod_{\alpha}
e^{t_{\alpha}(r_{\alpha})} \prod_{\beta} e^{t_{\beta}(r_{\beta})},
\label{eq:ben_psitrial}
\end{equation}
where we use for the binding parameters (in atomic units) $a=0.05$,
$c=0.06$.  The product over $\alpha$ and $\beta$ runs over the carbon
atoms and hydrogen atoms, respectively.  The atom-atom terms
$t_{\alpha}(r_{\alpha})$ and $t_{\beta}(r_{\beta})$ are functions of
$^4$He-carbon and $^4$He-hydrogen distances $r_{\alpha}$ and
$r_{\beta}$ respectively, and their analytical forms are chosen to
cancel out the leading singularities in the atom-atom potential energy
terms.\cite{mushinski94} In this study we use $t_{\alpha}(r_{\alpha})
= -c_{\alpha}r_{\alpha}^{-6}$, $t_{\beta}(r_{\beta}) =
-c_{\beta}r_{\beta}^{-5}$, with the parameters (in atomic units)
$c_{\alpha}=6000$, $c_{\beta}=8000$.  The trial function of
Eq.~(\ref{eq:ben_psitrial}) possesses the same $D_{6h}$ symmetry as
the $^4$He-benzene potential.

A ground state DMC calculation using the trial function and potential
discussed above gives a ground state energy $E_0=-21.61(2)$~cm$^{-1}$,
which corresponds to about 32\% of the global energy minimum of the
$^4$He-benzene potential.  Such a high zero-point energy is typical of
helium van der Waals systems,\cite{whaley98} and underscores the need
for a fully quantum mechanical treatment of the van der Waals degrees
of freedom.

\begin{table}[htb]
\caption[]{Operators $\hat{A}^{(\Gamma)}$ and the resulting energies
$E-E_0$ (in~cm$^{-1}$) for $^4$He-benzene van der Waals excitations.
For the two-dimensional irreducible representations, the two
projectors\index{POITSE!projector operators} listed give degenerate
energies.  The three rightmost columns list energies obtained from
hybrid branching/weighting~(B/W),\index{DMC!hybrid
branching/weighting} pure weighting~(PW),\index{DMC!pure weighting}
and pure branching~(PB).\index{DMC!pure branching}}
\begin{center}
\begin{tabular}{|c|c|c|c|c|}
\hline
$\Gamma$ &	$\hat{A}^{(\Gamma)}$ &	\multicolumn{3}{c|}{$E-E_0$} \\
\cline{3-5}
&		&			B/W &	PW &	PB \\
\hline
$E_{1g}$ &	$xz, yz$ &		6.39 &	6.39 &	6.39 \\
$E_{1u}$ &	$x, y$ &		7.04 &	6.97 &	7.04 \\
$A_{2u}$ &	$z(x^2+y^2)$ &		7.76 &	7.64 &	7.86 \\
$A_{1g}$ &	$x^2+y^2$ &		8.44 &	8.54 &	8.44 \\
$E_{2u}$ &	$z(x^2-y^2), xyz$ &	9.41 &	9.36 &	9.48 \\
$E_{2g}$ &	$x^2-y^2, xy$ &		9.96 &	9.84 &	10.01 \\
$B_{2u}$ &	$x^3-3xy^2$ &		11.22 &	11.34 &	11.19 \\
$B_{1g}$ &	$z(x^3-3xy^2)$ &	11.41 &	11.56 &	11.58 \\
$B_{2g}$ &	$z(y^3-3x^2y)$ &	13.34 &	13.39 &	13.25 \\
$B_{1u}$ &	$y^3-3x^2y$ &		13.58 &	13.58 &	13.37 \\
\hline
\end{tabular}
\end{center}
\label{table:ben_proj}
\end{table}
We choose the excitation operators\index{POITSE!projector operators}
$\hat{A}^{(\Gamma)}$ based on symmetry considerations, where the
superscript $\Gamma$ denotes an irreducible representation of the
$D_{6h}$ point group.  Since the trial function $\Psi_T({\bf r})$
transforms as the totally symmetric representation $A_{1g}$, for a
given $\hat{A}^{(\Gamma)}$, the integral
$\bra{\Psi_T}\hat{A}^{(\Gamma)}\ket{\phi_n}$ in Eq.~(\ref{eq:corr2d})
is only nonzero for states $\ket{\phi_n}$ which transform as $\Gamma$.
Thus an appropriate choice of $\hat{A}^{(\Gamma)}$ will, by symmetry,
significantly reduce the number of terms in the summation of
Eq.~(\ref{eq:corr2d}), leaving only decay terms whose characteristic
decay times are presumably more well-separated, and thus easier to
resolve.  The various choices of the operators $\hat{A}^{(\Gamma)}$ we
use here are listed in Table~\ref{table:ben_proj}, where
$\hat{A}^({\Gamma})$ is defined with respect to the benzene principal
axis frame centered on the benzene center-of-mass.  In this coordinate
system, the $x$-axis is perpendicular to the benzene C-C bond, the
$y$-axis lies along the benzene C-H bond, and the $z$-axis is
perpendicular to the benzene plane.

To evaluate the correlation function $\ktau$, we sample an initial
ensemble of 1000 walkers from every 100 steps of a VMC
walk.\index{VMC} This initial ensemble is propagated by a DMC sidewalk
with a time step of $\Delta\tau = 10$~Hartree$^{-1}$.  In the
$^4$He-benzene system, the energy differences of interest are
sufficiently large such that we can employ and compare all three DMC
implementations discussed in Sec.~\ref{subsec:implement}.  For the
hybrid branching/weighting scheme,\index{DMC!hybrid
branching/weighting} the ensemble size and sum of weights in the DMC
propagation are kept at approximately 1000 on average by choosing an
appropriate set of DMC parameters $w_{\mathit{min}},
w_{\mathit{max}}$, and $\eta$ (Eq.~(\ref{eq:eref})).  For DMC with
pure weights\index{DMC!pure weights} and DMC with pure
branching\index{DMC!pure branching}, the only adjustable parameter is
the update parameter $\eta$.  About 500 independent decays $\ktau$ are
generated in this manner, and subsequently used as input for the MEM
inversion, resulting in the spectral function $\kappa(E)$.  Each
choice of projector $\hat{A}^{(\Gamma)}$ results in a single dominant
peak in the corresponding $\kappa(E)$, and the peak positions are
listed in Table~\ref{table:ben_proj}.  These excited state energies
show general agreement (to within $\sim 0.2$~cm$^{-1}$) between the
three DMC implementations.

\begin{figure}[hbt]
\begin{center}
\scalebox{0.4}{\includegraphics{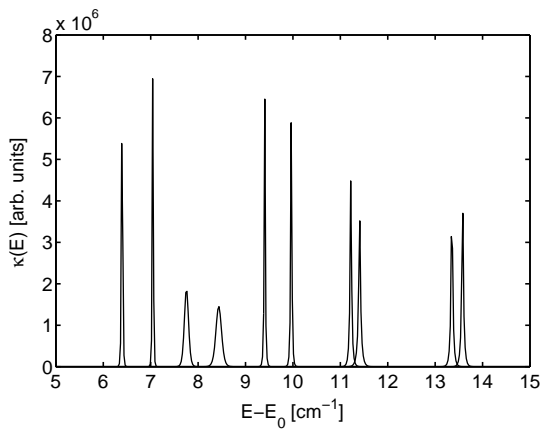}}
\end{center}
\caption[]{Spectral function $\kappa(E)$ for $^4$He-benzene, computed
using a hybrid branching/weighting approach.  Note that this plot
represents a superposition of $\kappa(E)$'s obtained from multiple
projectors, each yielding a single peak from the MEM inversion.}
\label{fig:ben_spectra}
\end{figure}
In Fig.~\ref{fig:ben_spectra} we superimpose the spectral functions
obtained using the hybrid branching/weighting approach.  There, the
peaks are grouped in doublets whose splittings range from $\sim
0.2-0.7$~cm$^{-1}$.  These doublets are due to projectors which are
symmetric and antisymmetric with respect to reflection about the
benzene plane.  They constitute a tunneling
splitting\index{tunneling!$^4$He-benzene} between the two equivalent
global potential minima along the benzene $C_6$-axis, above and below
the aromatic ring plane.  Tunneling of helium around a planar moiety
has also been observed in basis set calculations for the
2,3-dimethylnaphthalene$\cdot$He complex, where the magnitude of the
splittings ranged from $<10^{-4}$~cm$^{-1}$ for localized states up to
3.2~cm$^{-1}$ for highly delocalized states.\cite{bach97} The
tunneling splittings which we obtain here exhibit a decrease in
magnitude with increasing energy.  Since the energies of highest
levels correspond to about 12\% of the $^4$He-benzene potential energy
minimum, this decrease in the tunneling splitting can be attributed to
increasing anharmonicities in the $^4$He-benzene interaction potential
as these levels approach dissociation.  Inclusion of the benzene
rotational kinetic energy term into the Hamiltonian of
Eq.~(\ref{eq:ben_ham}) qualitatively changes the features of the
energy spectrum, removing this decrease in the tunnel splitting.  The
specific effects of this rotational contribution, as well as the
general physics of $^4$He$_N$-benzene clusters, will be reported in a
future study.\cite{huang01b}

\section{Conclusion}
We have extended the applicability of the POITSE method by introducing
branching processes in the DMC evaluation of an imaginary time
correlation function $\ktau$.  The effects of branching were tested in
the determination of excited state energies for two simple systems,
namely the one-dimensional ammonia inversion mode and the
six-dimensional $^4$He-benzene van der Waals modes.  While in an ideal
situation one would employ a pure weighting scheme in the DMC
propagation, in the ammonia study we were faced with the problem of
evaluating a slowly decaying $\ktau$ corresponding to a small
tunneling splitting.  Thus, the incorporation of branching in the DMC
sidewalk is essential for the stable computation of small energy
differences.  We have also provided a comparison between the various
branching schemes and the pure weighting scheme in the $^4$He-benzene
example, and have demonstrated that the results obtained are in good
numerical agreement.

The incorporation of branching as described here has been critical in
allowing excited state energies to now be evaluated for much larger
systems using the POITSE approach.\cite{huang01b,viel01b} Another
current modification in progress includes the implementation of
descendant weighting\index{DMC!descendant weighting}
techniques\cite{liu74,barnett91,casulleras95,hornik00} to construct an
estimate of the exact ground state wave function $\ket{\phi_0}$
``on-the-fly''.  The projector $\hat{A}$ would then operate on the
exact $\ket{\phi_0}$, instead of an approximate trial function
$\ket{\Psi_T}$.  These improvements in the general POITSE methodology
open the way for efficient and accurate Monte Carlo evaluation of
excited state energies for large systems.

\section*{Acknowledgments}
Financial and computational support from the National Science
Foundation through grant CHE-9616615.  An allocation of supercomputing
time from the National Partnership for Advanced Computational
Infrastructure (NPACI) is gratefully acknowledged.

\end{document}